\documentclass[%
 reprint,
 amsmath,amssymb,
 aps,
]{revtex4-2}

\usepackage{graphicx}
\usepackage{dcolumn}
\usepackage{bm}

\begin{document}


\title{Singlet cross section and the tensor interaction in $\overline{\mathrm{p}} \mathrm{p} \rightarrow \bar{\Lambda} \Lambda$ }

\author{Deepak Pachattu}
\affiliation{Department of Physics, BITS Pilani K K Birla Goa Campus, Zuarinagar 403726, Goa, India}
\begin{abstract}
    In this short paper, we demonstrate using irreducible tensorial techniques and in a model-independent way, why tensor interactions and also orbital-angular momentum-changing vector interactions are absent in the singlet unpolarized differential cross section for the reaction $\overline{\mathrm{p}} \mathrm{p} \rightarrow \bar{\Lambda} \Lambda$.
\end{abstract}
\maketitle
It is now well known that in the annihilation reaction, $\overline{\mathrm{p}} \mathrm{p} \rightarrow \bar{\Lambda} \Lambda$, the singlet contribution to the $\bar{\Lambda}\Lambda$ in the final state is suppressed (see \cite{tabakin} and references therein). The tensor interaction is supposed to play a dominant role in the $\bar{p}p$ $I=0$ channel and the reaction  $\overline{\mathrm{p}} \mathrm{p} \rightarrow \bar{\Lambda} \Lambda$ was considered a good place to study the effects of this force. In this short paper, we study this reaction a bit more closely, in a model-indpendent way using irreducible tensorial techniques especially since this reaction becomes relevant in the context of future experimental setups like PANDA \cite{panda}. In doing so, we find that the unpolarized singlet cross section does not get any contribution from the tensor interaction and also orbital angular momentum-changing vector interactions. To do this, we use the irreducible tensorial techniques developed earlier by us in \cite{grpnd}.

The irreducible formalism employed is identical to the one used for pion-production in $NN$ collisions \cite{grpnd}, with the momenta and angular quantum numbers associated with the pion set equal to zero. 
The matrix $\mathbf{M}$ in spin-space for $\overline{\mathrm{p}} \mathrm{p} \rightarrow \bar{\Lambda} \Lambda$ may be expressed in terms of the initial and final state-channel spins, $s_i,s_f=0,1$ as
\begin{equation}
\mathbf{M}=\sum_{s_f, s_i=0}^1 \sum_{\lambda=\left|s_f-s_i\right|}^{s_f+s_i}\left(S^\lambda\left(s_f, s_i\right) \cdot M^\lambda\left(s_f, s_i\right)\right),
\end{equation}
with the irreducible reaction amplitudes in channel-spin space given by 
\begin{eqnarray}
M_\mu^\lambda\left(s_f, s_i\right)= \sum_{l_f,l_i,j} (-1)^{l_i+s_i+l_f-j}[j]^2[s_f]^{-1}\\\nonumber 
\times W(l_fl_is_fs_i;\lambda j)
M_{l_f s_f; l_i s_i}^j(E)\left(Y_{l_f}\left(\hat{\mathbf{p}}_f\right) \otimes Y_{l_i}\left(\hat{\mathbf{p}}_i\right)\right)_\mu^\lambda,
\end{eqnarray}
in terms of the partial-wave amplitudes $M_{l_f s_f; l_i s_i}^j(E)$, which are functions of the c. m. energy $E$. The notations here follow \cite{grpnd}. 

If $\rho^i$ is the initial state density matrix, the density matrix $\rho^f $ characterizing the final spin-state is
$
\rho^f=\mathbf{M} \rho^i \mathbf{M}^{\dagger}
$,
which can be recast in terms of the irreducible Fano statistical tensors $t^k_q$ as 
\begin{equation}
\rho^f=\sum_{s_f, s_f^{\prime},k}\left(S^{k}\left(s_f, s_f^{\prime}\right) \cdot t^{k}\left(s_f, s_f^{\prime}\right)\right),
\end{equation}
with
\begin{eqnarray}
t_{q}^{k}\left(s_f, s_f^{\prime}\right)=\sum_{s_i, s_i^{\prime}} \sum_{\lambda, \lambda^{\prime}, \lambda^{\prime \prime}} \sum_{k’, k^{\prime\prime}}(-1)^{s_i^{\prime}-s_f^{\prime}}\left[s_f^{\prime}\right]\left[s_i\right][\lambda]\left[\lambda^{\prime}\right]\\\nonumber
\times
[k^{\prime\prime}]\left[k^{\prime}\right]^2\left[\lambda^{\prime \prime}\right](-1)^{\lambda+k^{\prime\prime}-k^{\prime}} W\left(s_i^{\prime} k^{\prime\prime} s_f \lambda ; s_i k^{\prime}\right)\\\nonumber
\times W\left(s_f^{\prime} \lambda^{\prime} s_f k^{\prime} ; s_i^{\prime} k\right) W\left(k^{\prime \prime} \lambda k  \lambda^{\prime} ; k^{\prime} \lambda^{\prime\prime}\right)\\\nonumber
\times\left(P^{k^{\prime\prime}}\left(s_i, s_i^{\prime}\right) \otimes\left(M^\lambda\left(s_f, s_i\right) \otimes M^{\dagger^{\lambda^{\prime}}}\left(s_f^{\prime}, s_i^{\prime}\right)\right)^{\lambda^{\prime \prime}}\right)_{q}^{k}
\end{eqnarray}
and the channel-spin polarization tensors 
\begin{eqnarray}
P_q^k\left(s_i, s_i^{\prime}\right)=\frac{1}{2}\left[s_i^{\prime}\right] \sum_{k_1, k_2}(-1)^{k_1+k_2-k}\left[k_1\right]\left[k_2\right]\\\nonumber
\times \left\{\begin{array}{ccc}
\frac{1}{2} & \frac{1}{2} & s_i \\
\frac{1}{2} & \frac{1}{2} & s_i^{\prime} \\
k_1 & k_2 & k
\end{array}\right\}\left(P^{k_1} \otimes Q^{k_2}\right)_q^k,
\end{eqnarray}
in terms of the spherical components
\begin{equation}
\begin{aligned}
P_0^0 & =Q_0^0=1 ; P_0^1=P_z ; Q_0^1=Q_z \\
P_{ \pm 1}^1 & =\mp \frac{1}{\sqrt{2}}\left(P_x \pm \mathrm{i} P_y\right) ; Q_{ \pm 1}^1=\mp \frac{1}{\sqrt{2}}\left(Q_x \pm \mathrm{i} Q_y\right)
\end{aligned}
\end{equation}
of the beam and target polarizations $\overrightarrow{P}, \overrightarrow{Q}$ respectively.
If the beam and target are unpolarized as it is in the case of \cite{tabakin,panda}, $P_q^k\left(s_i, s_i^{\prime}\right)=1/4$.
From eqs. (3) and (4) we can immediately see that the singlet and triplet contributions to the differential cross section are respectively $t^0_0(0,0)$ and $t^0_0(1,1)$, so that the differential cross section is 
\begin{equation}
\frac{\mathrm{d}\sigma}{\mathrm{d}\Omega}=\operatorname{Tr} \rho^f=\sum_{s_f=0,1}\left(2 s_f+1\right) t_0^0\left(s_f, s_f\right).
\end{equation}
If we use the channel-spin version of the projection-cum-spin flip operator introduced in \cite{grmsv}:
\begin{eqnarray}
S_{{\textstyle{\frac{1}{2}}},{\textstyle{\frac{1}{2}}}}\left(l_f s_f ; j ; l_i s_i\right)=\sum_\lambda (-1)^{1+l_f-j}\frac{[j]^2}{[s_f][l_f]}\\\nonumber
\times W(l_il_fs_is_f;\lambda j)\left(S^\lambda\left(l_f, l_i\right) \cdot S^\lambda\left(s_f, s_i\right)\right),
\end{eqnarray}
using which eq. (1) can now be rewritten as 
\begin{equation}
\mathbf{M}=\sum_{l_f, s_f, j, l_i, s_i} M_{l_f s_f ; l_i s_i}^j S_{{\textstyle{\frac{1}{2}}}, {\textstyle{\frac{1}{2}}}}\left(l_f s_f ; j ; l_i s_i\right) .
\end{equation}
We can now in general define an effective interaction \cite{grmsv} for this reaction through
\begin{equation}
\begin{aligned}
\left\langle\mathbf{r}_f\left|V_{\mathrm{eff}}\right| \mathbf{r}_i\right\rangle=\sum_{s_f,s_i\lambda}S^\lambda(s_f,s_i) \cdot\left\langle\mathbf{r}_f\left|V^{\left(s_f s_i\right) \lambda}\right| \mathbf{r}_i\right\rangle,
\end{aligned}
\end{equation}
where 
\begin{eqnarray}
\left\langle\mathbf{r}_f\left|V_\mu^{\left(s_f s_i\right)\lambda}\right| \mathbf{r}_i\right\rangle=\sum_{l_f,l_i} \left\langle r_f\left|V_{l_f s_f ; l_i s_i}^j\right| r_i\right\rangle\\\nonumber\times\left\langle\hat{\mathbf{r}}_f\left|S_\mu^\lambda\left(l_f, l_i\right)\right| \hat{\mathbf{r}}_i\right\rangle,    
\end{eqnarray}
with $\left\langle r_f\left|V_{l_f s_f ; l_i s_i}^j\right| r_i\right\rangle$ being the radial non-local terms. The contribution from the non-central interactions can be investigated further if we write the channel spin operators in terms of the Pauli Spin operators $\sigma^\lambda
_\mu({\textstyle{\frac{1}{2}}},{\textstyle{\frac{1}{2}}})$ as \cite{grpnd}
\begin{eqnarray} 
S_\mu^\lambda\left(s_f, s_i\right)=\frac{1}{2} \sum_{\lambda_1, \lambda_2=0,1}\left\{\begin{array}{ccc}{\textstyle{\frac{1}{2}}}&{\textstyle{\frac{1}{2}}}& s_f \\ {\textstyle{\frac{1}{2}}}&{\textstyle{\frac{1}{2}}}&s_i \\ \lambda_1 & \lambda_2 &\lambda\end{array}\right\} \left[s_f\right]^2\left[s_i\right]\\\nonumber \times \left[\lambda_1\right]\left[\lambda_2\right](-1)^{s_i-1} \left(\sigma^{\lambda_1}\left({\textstyle{\frac{1}{2}}}, {\textstyle{\frac{1}{2}}}\right) \otimes \sigma^{\lambda_2}\left({\textstyle{\frac{1}{2}}}, {\textstyle{\frac{1}{2}}}\right)\right)_\mu^\lambda.\end{eqnarray}
We can now rewrite eq. (10) as \cite{grmsv}
\begin{eqnarray}    
\left\langle\mathbf{r}_f\left|V_{\mathrm{eff}}\right| \mathbf{r}_i\right\rangle=\sum_{\lambda_1, \lambda_2, \lambda}&\left(\sigma^{\lambda_1} ({\textstyle{\frac{1}{2}}},{\textstyle{\frac{1}{2}}})\otimes \sigma^{\lambda_2} ({\textstyle{\frac{1}{2}}},{\textstyle{\frac{1}{2}}})\right)^\lambda \cdot\\\nonumber
&\left\langle\mathbf{r}_f\left|V^{\left(\lambda_1 \lambda_2\right) \lambda})\right| \mathbf{r}_i\right\rangle,
\end{eqnarray}
where 
\begin{eqnarray}\left\langle\mathbf{r}_f\left|V_\mu^{\left(\lambda_1 \lambda_2\right) \lambda}\right| \mathbf{r}_i\right\rangle=\sum_{l_f, s_f, j, l_i, s_i} G_{\lambda_1 \lambda_2 \lambda}\left(l_f s_f ; j ; l_i s_i\right) \\\nonumber\times\left\langle r_f\left|V_{l_f s_f ; l_i s_i}^j\right| r_i\right\rangle\left\langle\hat{\mathbf{r}}_f\left|\mathcal{S}_\mu^\lambda\left(l_f, l_i\right)\right| \hat{\mathbf{r}}_i\right\rangle,\end{eqnarray}
and $G_{\lambda_1 \lambda_2 \lambda}\left(l_f s_f ; j ; l_i s_i\right)$ are geometrical factors \cite{grmsv}. In eq. (13), $\lambda_1=\lambda_2=\lambda=0$ corresponds to spin-independent central interactions, $\lambda_1=\lambda_2=1$ with $\lambda=1,2$ corresponds to spin-dependent central interactions, while $\lambda_1=0(1), \lambda_2=1(0)$ with $\lambda=1$ corresponds to spin-dependent non-central interactions. The spin-orbit interactions are realized when the choice $S^\lambda_\mu(l,l)=\tau^k_q(\mathbf{S})$ is made \cite{grmsv}.
The singlet cross section $t^0_0(0,0)$, with unpolarized beam and target, which is the observable of interest, is readily obtained by putting $s_f=s_f'=0$, and $k=q=k''=0$ in eqs. (4) and (5). This also means $s_i=s_i'$, $\lambda''=0$, which in turn implies $\lambda=\lambda'=k'=s_i$, giving us
\begin{equation}
    t^0_0(0,0)=\frac{1}{4}\left[\left|M^0_0(0,0\right|^2+\sum_\mu\left|M^1_\mu(0,1)\right|^2\right].
\end{equation}
This can be further simplified if we pick a coordinate system in which $\overrightarrow{p_i}$ is along the positive $z$-axis and $\overrightarrow{p_f}$ is in the $x$-$z$ (scattering) plane. Then from eq. (2) and using the fact that $Y_{l_i m_i}\left(\hat{p}_i\right)=\delta_{m_i 0} \sqrt{\frac{2 l_i+1}{4 \pi}}$ and
\begin{equation}
Y_{l-m}(\theta, \phi)=(-1)^{-m} e^{-2 i m \phi} Y_{l m}(\theta, \phi),
\end{equation}
with $\phi=0$ for $\overrightarrow{p_f}$, we have
\begin{equation}
M^\lambda_\mu(s_f,s_i)=(-1)^{\lambda-\mu}M^\lambda_{-\mu}(s_f,s_i),
\end{equation}
since parity is conserved:
\begin{equation}
(-1)^{l_i}\pi_p\pi_{\bar{p}}=(-1)^{l_f}\pi_\Lambda\pi_{\bar{\Lambda}},    
\end{equation}
with $\pi_p\pi_{\bar{p}}=\pi_\Lambda\pi_{\bar{\Lambda}}=-1$, implying 
\begin{equation}    
(-1)^{l_f}=(-1)^{l_i}.
\end{equation}
This now further reduces eq. (15) to
\begin{equation}
    t^0_0(0,0)=\frac{1}{4}\left[\left|M^0_0(0,0\right|^2+2\left|M^1_1(0,1)\right|^2\right].
    \end{equation}
We now see that since $\lambda=s_i=0,1$, the highest value of $\lambda$ that contributes to the singlet cross section, is one, which means that the usual tensor interaction which corresponds to $\lambda=2$ {\em does not} contribute to the singlet cross section at all. At the most, we can have {\em vector} interactions which can contribute to spin-dependent non-central interactions, in the singlet state. But we will see that even these non-central interactions {\em cannot} change the orbital angular momentum, in the following way: from eq. (8), when $\lambda=s_i=0$, we see that trivially, $l_i=l_f$. When $\lambda=s_i=1$, we see from eq. (8) again that $l_i$ and $l_f$ can be the same or they can differ at the most by one unit of angular momentum. But from eq. (19), we see that $l_i=l_f$. Thus our irreducible tensor formalism clearly leads us, in a {\em {model-independent}} way, to conclude that the singlet cross section cannot contain the usual tensor interactions and also cannot have orbital angular momentum-changing vector interactions. However, for the triplet state, $s_f=1$ and for $s_i=1$, we can certainly have tensor ($\lambda=2$) contributions to the unpolarized triplet differential cross section.
\begin{acknowledgments}
The author acknowledges with gratitude the financial support received under the DST-SERB grant, CRG/2021/000435.
\end{acknowledgments}

\end{document}